\documentclass{PoS}
\usepackage{bm}

\title{Update on extraction of transversity PDF \\ 
from inclusive di-hadron production}

\ShortTitle{Update on extraction of transversity}

\author{\speaker{Marco Radici}%\thanks{A footnote may follow.} 
\\
INFN Sezione di Pavia, via Bassi 6, I-27100 Pavia, Italy  
\\
E-mail: \email{marco.radici@pv.infn.it}}

\abstract{The transversity was recently extracted from data on the production of hadron pairs in semi-inclusive deep-inelastic scattering. This analysis can be conveniently performed in the framework of collinear factorization where the elementary mechanism is represented by the simple product of transversity and of a suitable chiral-odd function describing the fragmentation of a transversely polarized parton into a pair of hadrons inside the same current jet. The same elementary mechanism was predicted long ago to generate an asymmetry in the azimuthal distribution of the hadron pairs when they are produced in proton-proton collisions with one transversely polarized proton. Recently, the STAR Collaboration has observed this asymmetry. We analyze the impact of these data on our knowledge of transversity and we present its first preliminary extraction from a global fit of all data in hard processes with inclusive di-hadron production. }

\FullConference{XXV International Workshop on Deep-Inelastic Scattering and Related Subjects\\
		3-7 April 2017\\
		University of Birmingham, UK}

\begin{document}

\section{Introduction}
\label{sec:intro}

The transverse polarization parton distribution $h_1$ (transversity) is the least known of parton distribution functions (PDFs) because it is not diagonal in the helicity basis (in jargon, it is a chiral-odd function). As such, $h_1$ is suppressed in simple processes like inclusive deep-inelastic scattering (DIS). It can be measured only in processes with at least two hadrons: semi-inclusive DIS (SIDIS) or hadronic collisions. On the other hand, the chiral-odd transversity is very peculiar. There is no transversity for gluons in a spin-$\textstyle{\frac{1}{2}}$ hadron. Therefore, $h_1$ scales under QCD evolution as a pure non-singlet function. Its first Mellin moment, the so-called tensor charge, can be useful to search for effects from physics beyond the Standard Model (BSM)~\cite{Courtoy:2015haa}.

We have a limited knowledge of transversity. Relying on an extended-parton-model description of the so-called Collins effect, transversity was extracted for the first time by simultaneously fitting data from single-hadron SIDIS on transversely polarized targets and data on emissions of two almost back-to-back hadrons in $e^+ e^-$ annihilations (see Ref.~\cite{Anselmino:2015sxa} and references therein). Recently, the analysis was carried out using the factorization framework and evolution equations that are consistent with the fact that the Collins mechanism is directly sensitive to the intrinsic transverse momentum of partons~\cite{Kang:2015msa}. Present fixed-target SIDIS data allow to access only the valence components of $h_1$. Within the (large) errors, both analyses give basically overlapping results~\cite{Kang:2015msa}. 

Alternatively, transversity can be extracted in the standard framework of collinear factorization using data from two-hadron SIDIS. In fact, the related cross section at leading twist contains a term which is proportional to the product of $h_1$ and $H_1^{\sphericalangle}$, a specific chiral-odd Di-hadron Fragmentation Function (DiFF)~\cite{Radici:2001na,Bacchetta:2002ux}. The same $H_1^{\sphericalangle}$ appears in the leading-twist cross section for the semi-inclusive emission of back-to-back hadrons pairs in $e^+ e^-$ annihilations~\cite{Boer:2003ya}. The $H_1^{\sphericalangle}$ was parametrized from the $e^+ e^-$ data of {\tt BELLE} for production of $(\pi^+ \pi^-)$ pairs~\cite{Courtoy:2012ry}. The valence components $h_1^{u_v}$ and $h_1^{d_v}$ of transversity were subsequently extracted from the {\tt HERMES} and {\tt COMPASS} data for SIDIS production of $(\pi^+ \pi^-)$ pairs~\cite{Bacchetta:2011ip,Bacchetta:2012ty}. Recently, the analysis was updated by enclosing the latest and more precise {\tt COMPASS} data for a transversely polarized proton target~\cite{Radici:2015mwa}. The fit is based on a functional form for $h_1$ that satisfies the Soffer bound at any scale, and whose low-$x$ behaviour is constrained to give a finite integrated tensor charge; in fact, present fixed-target data do not put sufficient bounds on the low-$x$ trend. 
The functional form was taken enough flexible to accommodate different scenarios with one, two, or three nodes, and varying as well the normalization of the strong coupling constant at the $Z$ boson mass to account for the theoretical uncertainty in the determination of the $\Lambda_{\rm QCD}$ parameter. The outcomes turned out to be rather stable against these variations~\cite{Radici:2015mwa}. In the following, only the results from the scenario with two nodes (the so-called {\it flexible} scenario) will be shown. 
The error analysis was carried out by fitting 100 different replicas of data obtained by altering the experimental points with a random noise within the experimental uncertainty~\cite{Radici:2015mwa}. 
This approach is more general than the traditional Hessian method. It does not rely on approximating the parameter dependence by a linear expansion around the minimum; moreover, it gives a more realistic description of the statistical uncertainty, particularly when extrapolation outside the range of experimental data is needed~\cite{Bacchetta:2012ty}. 

A general consistency is observed between single-hadron and two-hadron SIDIS extractions, at least for $0.0065 \leq x \leq 0.34$ where the SIDIS data are~\cite{Kang:2015msa}, except for the valence down quark. For $h_1^{d_v}$ at $x \gtrsim 0.1$ and $Q^2 = 2.4$ GeV$^2$, all replicas of the transversity fitting two-hadron SIDIS tend to saturate the lower limit of the Soffer bound, forming a very narrow band clearly separated from the results from the Collins effect~\cite{Radici:2015mwa}. 
This trend is visible in all explored scenarios, indicating that it is not an artifact of the chosen functional form. Rather, it is 
%This trend is 
driven by some of the fitted {\tt COMPASS} data. In fact, we verified that by excluding from the fit the bins n.7 and n. 8 of the data set the deuteron target, the resulting replicas do not prematurely saturate the lower Soffer bound: they show a higher degree of flexibility by fluctuating within the lower and upper Soffer bounds, thus displaying a better degree of compatibility with the extraction based on the Collins effect~\cite{Radici:2016opu}. 

The main advantage of extracting the transversity by using data for inclusive di-hadron production lies in the possibility of working in a collinear factorization framework. Thus, transversity can be studied also in hadronic collisions and we can test the universality of the elementary mechanism $h_1 \, H_1^{\sphericalangle}$, as it was predicted long ago in Ref.~\cite{Bacchetta:2004it}. This is a unique opportunity with respect to the extraction based on the Collins effect, where factorization is broken because it is not possible to determine two well separated hard and soft scales. 
%because there are explicit counterexamples showing that the factorization is broken for single-hadron production in hadronic collisions where transverse momenta are not integrated~\cite{Rogers:2010dm}, the reason being related to the lack of two well separated hard and soft scales.  

Recently, experimental evidence for the predicted azimuthally asymmetric distribution of $(\pi^+ \pi^-)$ pairs in the $p p^\uparrow \rightarrow \pi^+ \pi^- X$ process, has been released by the {\tt STAR} Collaboration. Using the $h_1$ and $H_1^{\sphericalangle}$ obtained from the fit to the SIDIS and $e^+ e^-$ di-hadron data, we have compared our predictions based on the $h_1 \, H_1^{\sphericalangle}$ mechanism with the {\tt STAR} data, obtaining an asymmetry with size and shape in very reasonable agreement with the experimental measurements~\cite{Radici:2016lam}. This is an important achievement because it suggests that the same universal elementary mechanism is active in all hard processes leading to the inclusive production of $(\pi^+ \pi^-)$ pairs, thus opening the way to include also hadronic collision data for the extraction of transversity. 
Incidentally, by excluding from the fit the above mentioned deuteron bins from the {\tt COMPASS} SIDIS data set we also notice that discrepancies at forward rapidities between theoretical predictions and {\tt STAR} data get reduced~\cite{Radici:2016opu}. In turn, this suggests that proton-proton collision data would have a large impact on our knowledge of transversity if included in a global fit. Therefore, such a global fit is timely. In the following, we will show some preliminary results along this line.

%%%%%%%%%%%%%%%%%%%%%%%%%

\section{Formalism}
\label{sec:theory}

Data for inclusive di-hadron production are available for SIDIS, electron-positron annihilation, and proton-proton collisions. A global fit needs to address the relevant observables for each of these hard processes. 
%Here below, we review the main formulae for the relevant observables to be used in a global fit involving data for all hard reactions. 

As for SIDIS, at leading twist the cross section contains an azimuthally asymmetric term which is related to the elementary mechanism $h_1 \, H_1^{\sphericalangle}$. The corresponding single-spin asymmetry is described in detail in Refs.~\cite{Bacchetta:2011ip,Bacchetta:2012ty, Radici:2015mwa}. 
The DiFFs can be extracted from the electron-positron annihilation process leading to two correlated hadron pairs in opposite hemispheres. Again, at leading twist the cross section contains an azimuthally asymmetric term from which one can construct the so-called Artru-Collins asymmetry~\cite{Boer:2003ya,Courtoy:2012ry}. 
%The DiFFs can be extracted from the annihilation process $e^+ e^- \to (H_1 H_2) (\bar{H}_1 \bar{H}_2) X$. Again, at leading twist the cross section contains an azimuthally asymmetric term modulated by $\cos (\phi_R + \bar{\phi}_R)$, where $\phi_R$ and $\bar{\phi}_R$ give the orientation of the planes containing the momenta of the hadron pairs with respect to the lepton plane defined by the directions of the positron momentum and of the axis $\hat{\bm{z}} = - \bm{P}_h$; the latter vectors form the relative angle $\theta_2$ (see Fig.~1 of Ref.~\cite{Courtoy:2012ry} for more details). After averaging over $\cos\theta$, from the coefficient of the asymmetric term we can defined the so-called Artru-Collins asymmetry~\cite{Boer:2003ya,Courtoy:2012ry}
%\begin{equation}
%A_{e+e-} (\theta_2, z, M_h, \bar{z}, \bar{M}_h, Q^2) = \frac{\sin^2 \theta_2}{1 + \cos^2 \theta_2}\,  \frac{|\bm{R}|}{M_h} \, \frac{|\bm{\bar{R}}|}{\bar{M}_h} \, \frac{\sum_q e_q^2\, H_1^{\sphericalangle\, q}(z,M_h; Q^2)\, H_1^{\sphericalangle\, \bar{q}}(\bar{z},\bar{M}_h; Q^2)}{\sum_q e_q^2\, D_1^q(z,M_h; Q^2)\, D_1^{\bar{q}}(\bar{z},\bar{M}_h; Q^2)} \, .
%\label{eq:e+e-ssa}
%\end{equation} 
The DiFFs can be extracted by summing upon one of the two hemispheres, involving the number densities of fragmenting (un)polarized quarks~\cite{Courtoy:2012ry}. 

Lastly, in proton-proton collisions with one transversely polarized proton, at leading order in the hard scale, the cross section has an azimuthally asymmetric term that leads to the spin asymmetry described in Refs.~\cite{Bacchetta:2004it,Radici:2016lam}. 
Despite working in the collinear framework, the elementary combination $h_1 \, H_1^{\sphericalangle}$ happens convoluted with other ingredients in the numerator of such asymmetry, making its repeated computation very demanding for a global fit. This is a well known problem in all attempts of extracting PDFs by fitting hadronic collision data~\cite{Stratmann:2001pb}. In order to speed up the execution of the fitting code and to make a global fit feasible, the parametric part of the integrand (in our case, the transversity) is rewritten in terms of its Mellin anti-transform:
%A global fit to all data for inclusive di-hadron production proceeds through the above formulae for $A_{\rm SIDIS}$, $A_{e+e-}$, and $A_{pp}$. However, despite the collinear framework the elementary combination $h_1^b \, H_1^{\sphericalangle c}$ happens convoluted with the elementary cross section $d\Delta \hat{\sigma}_{ab^\uparrow \to c^\uparrow d}$ in the spin asymmetry of Eq.~(\ref{eq:App}) for proton-proton collisions. This is a well known problem in all attempts of extracting PDFs by fitting hadronic collision data~\cite{Stratmann:2001pb}. In order to speed up the execution of the fitting code and to make a global fit feasible, the parametric PDFs in Eq.~(\ref{eq:App}) (in our case, the transversity $h_1^b$) are rewritten in terms of their Mellin anti-transform:
\begin{eqnarray}
& &A_{pp} (\eta, |\bm{P}_T|, M_h) \nonumber \\
& &\quad = \frac{2 \, |\bm{P}_T|\, |\bm{S}_{BT}|}{d\sigma^0} \, \frac{|\bm{R}_T|}{M_h}  \sum_{a,b,c,d} \int \frac{dx_a \, dx_b}{16 \pi z_h} \, f_1^a(x_a) \, \frac{1}{2\pi i} \int_{C_N} dN \, \frac{1}{x_b^N}\, h_1^b(N) \, \frac{d\Delta \hat{\sigma}_{ab^\uparrow \to c^\uparrow d}}{d\hat{t}} \, H_1^{\sphericalangle c}(z_h, M_h^2) \nonumber \\
& &\quad = \frac{1}{2\pi i} \sum_b \int_{C_N} dN \, h_1^b(N) \, \hat{A}_{pp}^b (\eta, |\bm{P}_T|, M_h, N) \, ,
\label{eq:AppMellin}
\end{eqnarray}
where $d\sigma^0$ is the unpolarized cross section. The elementary process is the annihilation of partons $a$ and $b$ (carrying fractional momenta $x_a$ and $x_b$, respectively) into partons $c$ and $d$. The $d\Delta\hat{\sigma}$ is the elementary cross section for the case when the parton $b$ is transversely polarized and is contained in the proton with transverse polarization $\bm{S}_{BT}$. The transversely polarized parton $c$ fragments into a hadron pair with invariant mass $M_h$, total momentum with transverse component $\bm{P}_T$ to the incoming beam, and relative transverse momentum $\bm{R}_T$. The $\eta$ is the pseudo-rapidity of the process, and the pair fractional energy $z_h$ is bounded by momentum conservation in the elementary annihilation~\cite{Bacchetta:2004it}. 
The $C_N$ is a contour in the complex-$N$ plane and the dependence on the hard scale $|\bm{P}_T|$ is understood in each PDF and DiFF. The "reduced" asymmetry $\hat{A}_{pp}$ can be pre-calculated and stored before starting any minimization procedure to constrain the parametric expression of $h_1^b$, thus considerably reducing the execution time of the fit. Moreover, a suitable choice of $C_N$ can improve the convergence of the integral, further reducing the cpu time~\cite{Stratmann:2001pb}. However, the parametric form of transversity currently used in the fit of di-hadron SIDIS and $e^+ e^-$ data prevents from analytically computing its Mellin transform $h_1^b (N)$. In the following section, we will present preliminary results for the extraction of transversity from a global fit, by making a choice of $C_N$ that represents a convenient compromise between the convergence of the integral upon $dN$ and the need of speeding up the computation of $A_{pp}$.

%%%%%%%%%%%%%%%%%%%%%%%%%

%%%%%%%%%% Fig. 1  %%%%%%%%%
\begin{figure}
\begin{center}
\includegraphics[width=0.45\textwidth]{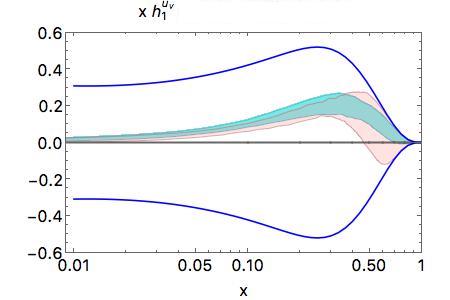} \hspace{0.1cm} 
\includegraphics[width=0.45\textwidth]{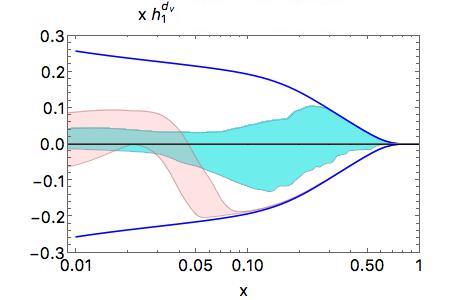}
\end{center}
\vspace{-0.5cm}
\caption{The transversity $x h_1 (x)$ as a function of $x$ at $Q^2 = 2.4$ GeV$^2$. Left panel for the valence up quark, right panel for the down quark. Dark (blue) lines represent the upper and lower Soffer bounds. Light (pink) band in background is the 68\% of 100 replicas from the latest Pavia fit of Ref.~\cite{Radici:2015mwa} using SIDIS and $e^+ e^-$ data; darker (cyan) band in foreground is the result of the 68\% of 200 replicas from the preliminary global fit here discussed, including also data from polarized p-p collisions.}
\label{fig:oldnew}
\end{figure}
\vspace{-0.5cm}
%%%%%

\section{Preliminary results of global fit}
\label{sec:results}

In the following pictures, each dark (cyan) uncertainty band in foregorund is formed by the central 68\% of 200 replicas of DiFFs and transversity that fit the data for inclusive di-hadron production in $e^+ e^-$ from {\tt BELLE}, SIDIS from {\tt HERMES} and {\tt COMPASS}, and proton-proton collisions from {\tt STAR}, after their parametric expression at $Q_0^2 = 1$ GeV$^2$ is evolved to the hard scale of each data bin using standard DGLAP evolution~\cite{Ceccopieri:2007ip}. 

In Fig.~\ref{fig:oldnew}, we show the transversity $x h_1 (x)$ as a function of $x$ at $Q^2 = 2.4$ GeV$^2$. In the left panel, the results refer to the valence up component, in the right panel to the valence down quark. The dark (blue) lines represent the upper and lower Soffer bounds. The light (pink) band in background is the uncertainty band from the latest upgrade of the Pavia fit in Ref.~\cite{Radici:2015mwa}, obtained by selecting the 68\% of 100 replicas extracted by fitting replicated data for $(\pi^+ \pi^-)$ pair inclusive production in SIDIS and $e^+ e^-$ annihilations taken by {\tt HERMES}, {\tt COMPASS}, and {\tt BELLE}, respectively. It is rather evident that including the {\tt STAR} data leads to a higher precision in our knowledge of the valence up component, while the anomalous squeezing of the replicas for the down component towards the saturation of the lower Soffer bound has disappeared. The dark (cyan) band in foreground in the right panel indicates that replicas now have much higher flexibility by fluctuating between the lower and higher Soffer bounds. This result suggests that hadronic collision data for inclusive di-hadron production can have a large impact on our knowledge of transversity. Their inclusion reduces the statistical weight of the peculiar bins n. 7, 8 from the {\tt COMPASS} SIDIS data set for a deuteron target. In fact, the net effect on the extracted transversity is similar to what we obtained by excluding those bins from the older fit of just SIDIS and $e^+ e^-$ data~\cite{Radici:2016opu}. The final outcome is a better degree of compatibility with the extraction of transversity from the Collins effect, as it results from the following figure.

\begin{figure}
\begin{center}
\includegraphics[width=0.45\textwidth]{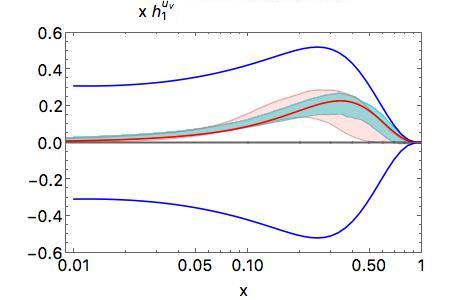} \hspace{0.1cm} 
\includegraphics[width=0.45\textwidth]{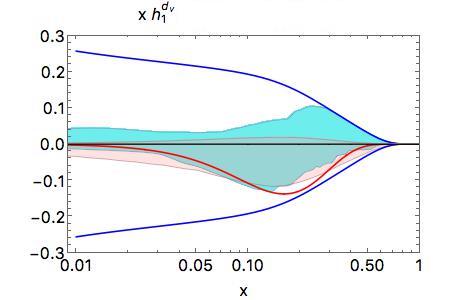}
\end{center}
\vspace{-0.5cm}
\caption{Same content and notation as in previous figure except for the light (pink) band in background, which is the result of the Torino fit of Ref.~\cite{Anselmino:2015sxa}, and the thick (red) line which is the average value obtained from the analysis of the Collins effect in the TMD factorization framework of Ref.~\cite{Kang:2015msa}.}
\label{fig:compare}
\end{figure}
%%%%%

In Fig.~\ref{fig:compare}, we show again the transversity $x h_1 (x)$ as a function of $x$ at $Q^2 = 2.4$ GeV$^2$ for the valence up quark (left panel) and the down quark (right panel). The darker (cyan) band in the foreground still represents the result of the current preliminary global fit. The lighter (pink) band is now the result of the Torino fit of Ref.~\cite{Anselmino:2015sxa} using the Collins effect. The central thick (red) line is the average most probable value obtained by Kang and collaborators in the analysis of the Collins effect within the TMD factorization framework~\cite{Kang:2015msa}. It is evident that the new (preliminary) global fit produces a transversity that is more compatible with the one extracted through the Collins effect, particularly for the down component. We note also that darker (cyan) band for the up quark is more narrow than the lighter (pink) one in the background. This higher precision is a direct benefit of having the possibility of enlarging the analysis to hadronic collision data, which is not possible in the single-hadron inclusive case because of the lack of a suitable factorization theorem.

%%%%%%%%%%%%%%%%%%%%%%%%%

\section{Conclusions}
\label{sec:end}

The transversity distribution can be reliably and uniquely extracted by fitting the combined set of data for the inclusive production of di-hadron pairs in $e^+ e^-$ annihilations, SIDIS, and proton-proton collisions, because the extraction is performed in the collinear factorization framework. 
It can be used as a cross-check of the extraction based on the Collins effect. But it uniquely allows to incorporate data from hadronic collisions, which is not possible for single-hadron inclusive production. 
The preliminary results of the global fit show a higher degree of compatibility with the extraction of transversity from the analysis of the Collins effect, with respect to the previous Pavia fits. In particular, the central 68\% of replicas for the valence down component do not show an anomalous premature saturation of the Soffer bound but display a higher flexibility. Moreover, the uncertainty band for the valence up component is more narrow. This better precision is important because it reflects in the corresponding tensor charge that is being considered in searches of new BSM effects. 

We stress again that the presented results are preliminary. Some ingredients of the analysis are still largely unconstrained, like, e.g., the gluon component of the unpolarized DiFF $D_1^g$. %, or need a deeper study, like the convergence of the Mellin integral in Eq.~(\ref{eq:AppMellin}). Work is in progress along these lines. 
Moreover, since the currently adopted parametric form of transversity cannot be analytically transformed in Mellin space, the contour $C_N$ in Eq.~(\ref{eq:AppMellin}) was chosen to trade the convergence of the integral with speeding up its computation to make the global fit feasible. In future developments, it may be necessary to change the adopted functional form of $h_1$ to simplify the calculations and match the computational time with the need of evaluating the integral in Eq.~(\ref{eq:AppMellin}) with sufficient precision to grant an overall convergence. 

%%%%%%%%%%%%%%%%%%%%%%%%%

\section*{Acknowledgments}

Most of the results presented in this report have been carried out in collaboration with A.~ Bacchetta, to whom I am deeply indebted. This research is partially supported by the European Research Council (ERC) under the European Union's Horizon 2020 research and innovation program (Grant Agreement No. 647981, 3DSPIN).

%%%%%%%%%%%%%%%%%%%%%%%%%

\bibliographystyle{JHEP}
\bibliography{mybiblio}

%\begin{thebibliography}{99}
%\bibitem{...}
%
%\end{thebibliography}

\end{document}